\documentclass[journal=jpclcd,manuscript=article]{achemso}
\usepackage[version=4]{mhchem} 
\usepackage{achemso}
\usepackage{amsfonts}
\usepackage{graphicx} 
\usepackage{dcolumn} 
\usepackage{bm} 
\usepackage[utf8]{inputenc}
\usepackage[T1]{fontenc}
\usepackage{etoolbox}
\usepackage{physics}
\usepackage{hyperref} 
\usepackage[version=4]{mhchem} 
\usepackage{xcolor}

\title{Spin-Dependent Stereochemistry: A Non-adiabatic Quantum Dynamics Case Study of \ce{S + H2 -> SH + H} Reaction} 

\author{Xuezhi Bian}
\email{xzbian@sas.upenn.edu}
\affiliation{Department of Chemistry, University of Pennsylvania, Philadelphia, Pennsylvania 19104, USA}
\author{Joseph E. Subotnik}
\email{subotnik@sas.upenn.edu}
\affiliation{Department of Chemistry, University of Pennsylvania, Philadelphia, Pennsylvania 19104, USA}
\date{\today}
 
\begin{document}
\begin{tocentry}
\includegraphics[width=5cm]{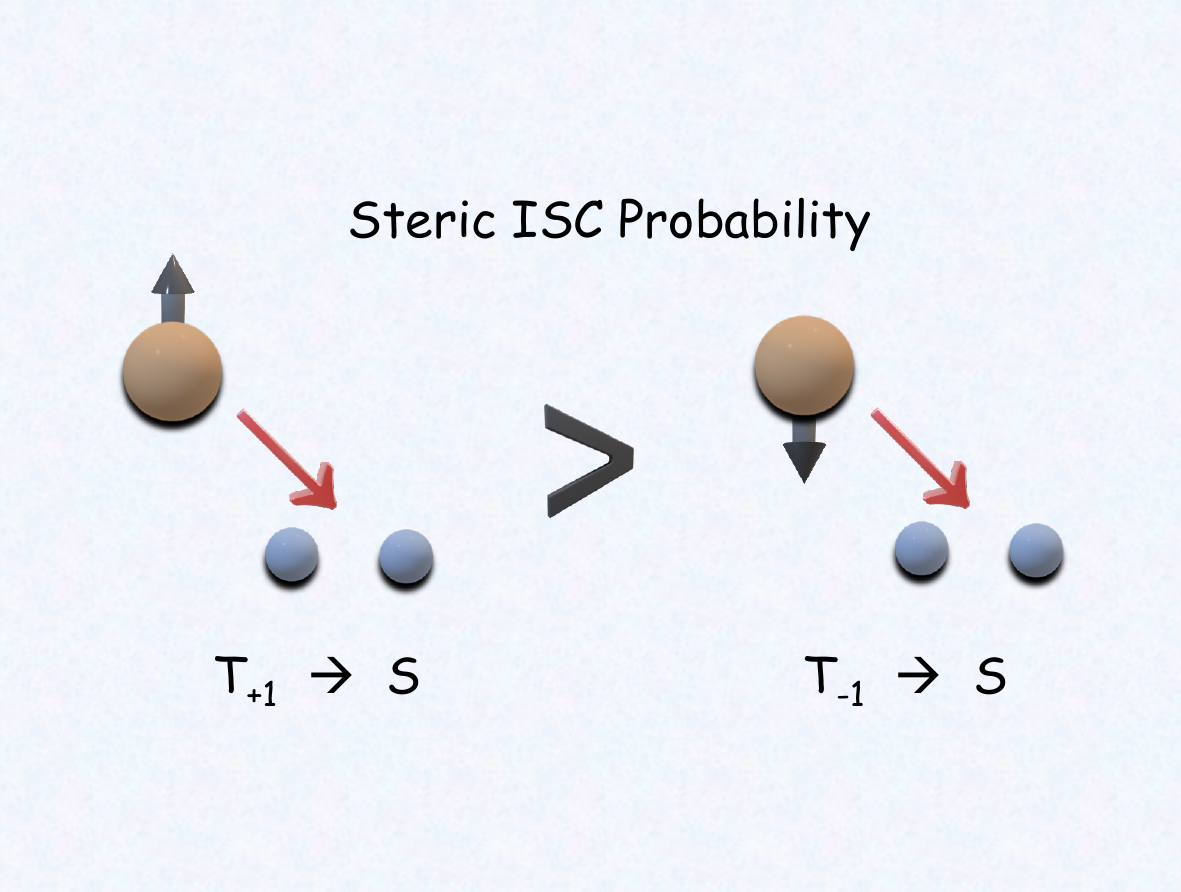}
\end{tocentry}

\begin{abstract}
We study the spin-dependent stereodynamics of the \ce{S + H2 -> SH + H} reaction using full-dimensional quantum dynamics calculations with zero total nuclear angular momentum along the triplet ${}^3A''$ states and singlet ${}^1A'$ states. 
We find that the interplay between the electronic spin direction and the molecular geometry has a measurable influence on the singlet-triplet intersystem crossing reaction probabilities.  
Our results show that for some incident scattering angles in the body-fixed frame, the relative difference in intersystem crossing reaction probabilities (as determined between spin up and spin down initial states) can be as large as $15\%$. 
Our findings are an {\em ab initio} demonstration of spin-dependent nonadiabatic dynamics which we hope will shine light as far as understanding the chiral-induced spin selectivity effect. 
\end{abstract}

\maketitle
\newpage

Coupled spin-nuclear-electronic dynamics are a very intriguing research field. In principle, chemists are coming to believe that they can control electronic spin polarization by running a current through a chiral molecule through the chiral induced spin selectivity (CISS) effect \cite{Ray1999,Gohler2011,Naaman2012,Naaman2019,Naaman2020,Eckvahl2023}. Vice versa, evidence is also emerging that one can  engineer chemical reactions by controlling the spin of reaction intermediates (either through CISS \cite{Mtangi2015,Ghosh2020,Vadakkayil2023} or simply applying a magnetic field \cite{Steiner1989}). In both scenarios, the interaction between (nuclear) molecular dynamics and  electronic spin play a pivotal role. However, a theoretical and computational understanding of these coupled dynamics from first principles is extremely challenging given the number of closely-spaced states that must be coupled together (correctly) in a beyond Born-Oppenheimer fashion.
For example, even for the simplest singlet-triplet intersystem crossing (ISC) model mediated by spin-orbit coupling (SOC), at least four electronic states are involved and the three triplets can couple differently to the singlet. 
While many useful nonadiabatic simulations of ISC have been performed to date, using both  wavepacket methods \cite{Hoffmann2000,Maiti2003,wang2022,Guan2023} and trajectory surface hopping methods \cite{Tully1990,Cui2014,Mai2015,Zaari2015}, very few studies have focused on disentangling the spin-related effects with nuclear dynamics and correlating the geometry of the spin state with chemical outcomes. In other words, for an ISC reaction, there is certainly literature pointing out that a singlet can convert into a triplet (and therefore change molecular geometry), but there is very little written about whether or not that transformation depends on which $M_s$ triplet state (of the  three possible triplet states $M_s = \left\{1,0,-1\right\}$) is produced. Moreover, looking forward, if one is to model such effects, it is essential to go beyond model systems and employ a realistic molecular potentials (ideally {\em ab initio} potentials) in order to understand practically how  spin-dependent nonadiabatic dynamics may or may not control chemical reactions \cite{Bian2021}.

One way to address the question above is to focus on stereodynamical scattering.  
Historically, chemists have recognized that nuclear (usually vibrational) degrees of freedom (DoF) can serve as an efficient way to control chemical reactions \cite{Schatz1984,Sinha1990,Guettler1994,Zare1998}, and recent experimental advances have focused on stereodynamics -- a new promising tool even for non-polar molecules \cite{Loesch1990,Aldegunde2005,Liu2016,Wang2023,Yang2023}. Within such gas phase scattering experiments, for example, it is known that if one  changes the molecular orientation, one can change the reactivity. For our purposes, we would like to investigate whether there is a connection between the final outcome of a reaction and (i) the nuclear orientation  and (ii) the direction of electronic spin polarization just before a reaction.

In this work, to address this question and explore the potential for spin-dependent stereochemistry, we will model the \ce{S + H2 -> SH + H} insertion reaction.  Given the sulfur atom ground state is a triplet state, we will initialize the incident \ce{S} atom with different spin polarizations and scan various scattering angles.
The \ce{S + H2 -> SH + H} reaction is an important experimental and theoretical prototype for studying kinetics and dynamics of gas phase bimolecular reactions \cite{Zhang2002,Maiti2004,Chu2007,Berteloite2010,Duan2013}, and previous studies (Ref.\citenum{Maiti2004}) have revealed significant intersystem crossing and nonadiabatic effects. 
Most importantly,  the system is relatively small, which will allow us (i) to utilize high level electronic structure theories (here, multireference configuration interaction (MRCI)) so as to generate accurate global potential energy surfaces (PESs); and (ii) to propagate exact full-dimensional nuclear quantum dynamics on these surfaces by wavepacket method. 
Overall, the \ce{S + H2 -> SH + H} system presents an ideal platform for investigating spin-dependent stereodynamics, which is the goal of this letter.

\section{Methods}

Based on the arguments of Hoffmann, Maiti and Schatz\cite{Hoffmann2000,Maiti2003,Maiti2004}, the most important low-energy intersystem crossings for the \ce{SHH} system are between the singlet  ${}^1   A'$ symmetry (where the isolated sulfur atom is \ce{S({}^1D)}) and two triplets  are of  ${}^3  A''$ and  ${}^3  A'$ symmetry (and the isolate sulfur atom is \ce{S({}^3P)}). By factorizing the electronic wavefunction into spatial and spin components and applying symmetry arguments, one can reduce this seven-state basis to four states if one cares only about the singlet-triplet ISC dynamics: 
only three (out of the possible six) triplet states can be coupled directly to the singlet by SOC, namely $\ket{{}^3{ A}''(M_s = \pm 1)}$ 
and $\ket{{}^3{ A}'(M_s = 0)}$ where we define the $z$-direction to be the direction perpendicular to the triatomic plane. Moreover, in this article, we are primarily interested in the $\ket{{}^3{ A}''(M_s = \pm 1)}$ states  since the $\ket{{}^3{ A}'(M_s = 0)}$ state is time-reversal symmetric (and so studying this state cannot not yield information about possible spin polarization). Therefore, we further simply the model by considering only the two triplet states ($\ket{{}^3{ A}''(M_s = \pm 1)}$ ) and the singlet state ${}^1 A'$.

Within a Born-Oppenheimer framework, the three-state intersystem crossing Hamiltonian is given by:
\begin{equation} \label{eq:Htot}
    \hat H_{\rm tot} = \hat T_{\rm n} \mathbb{I}_3 + \hat H_{\rm ISC}, 
\end{equation}
\begin{equation}\label{eq:ISC}
    \hat H_{\rm ISC} = 
    \begin{pmatrix}
    E_{\rm S} & V & -V^* \\ 
    V^* & E_{\rm T} & 0\\ 
    -V & 0  & E_{\rm T} 
    \end{pmatrix}.
\end{equation}
Here, $T_{\rm n}$ is the nuclear kinetic operator, $ \mathbb{I}_3$ is a $3\times 3$ identity matrix,  $E_{\rm S}$ and $E_{\rm T}$ are the energies of the ${}^1 A'$ and ${}^3 A''$ states respectively. $V = \bra{{}^1{ A'}} H_{\rm SO} \ket{{}^3{ A}''(M_s = +1)}$ represents the SOC matrix element between ${}^1 A'$ and ${}^3 A''(M_s = +1)$. 

\begin{figure}[ht]
    \centering
    \includegraphics[scale=0.75]{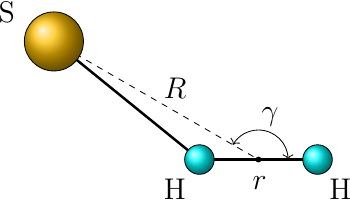}
    \caption{The Jacobi coordinates used in this paper for the \ce{S + H2} system. The \ce{S} atom is scattered towards \ce{H2} molecule with different Jacobi angles $\gamma$ and initial spin polarization.}
    \label{fig:coord}
\end{figure}

As far defining our reduced space of coordinates, we initialized our system with zero total nuclear angular momentum $\bm J_{nuc} = 0$. 
The system is then described in the body-fixed (BF) frame by three reactant-based Jacobi coordinates ($R,r,\gamma$) as shown in Fig.~{\ref{fig:coord}}.   
In such coordinates, the nuclear kinetic operator is of the form:
\begin{equation}\label{eq:Tn}
   \hat T_{\rm n} = - \frac {\hbar^2} {2\mu_R}\frac {\partial^2}{\partial R^2}  -  \frac {\hbar^2} {2\mu_r}\frac {\partial^2}{\partial r^2} + \frac {{\hat j}^2} {2I}  ,
\end{equation}
where $\mu_R = \frac {m_{\rm S} (m_{\rm H}+m_{\rm H})} {(m_{\rm S}+m_{\rm H}+m_{\rm H})}$ and $\mu_r = \frac {m_{\rm H}m_{\rm H}}{m_{\rm H}+m_{\rm H}}$ are the reduced masses for Jacobi coordinates $R$ and $r$,
$\hat j$ is the rotational angular momentum operator for $\ce{H2}$ molecule and $I = (1/\mu_RR^2 + 1/\mu_rr^2)^{-1}$.

\begin{figure}[ht]
    \centering
    \includegraphics[width=\linewidth]{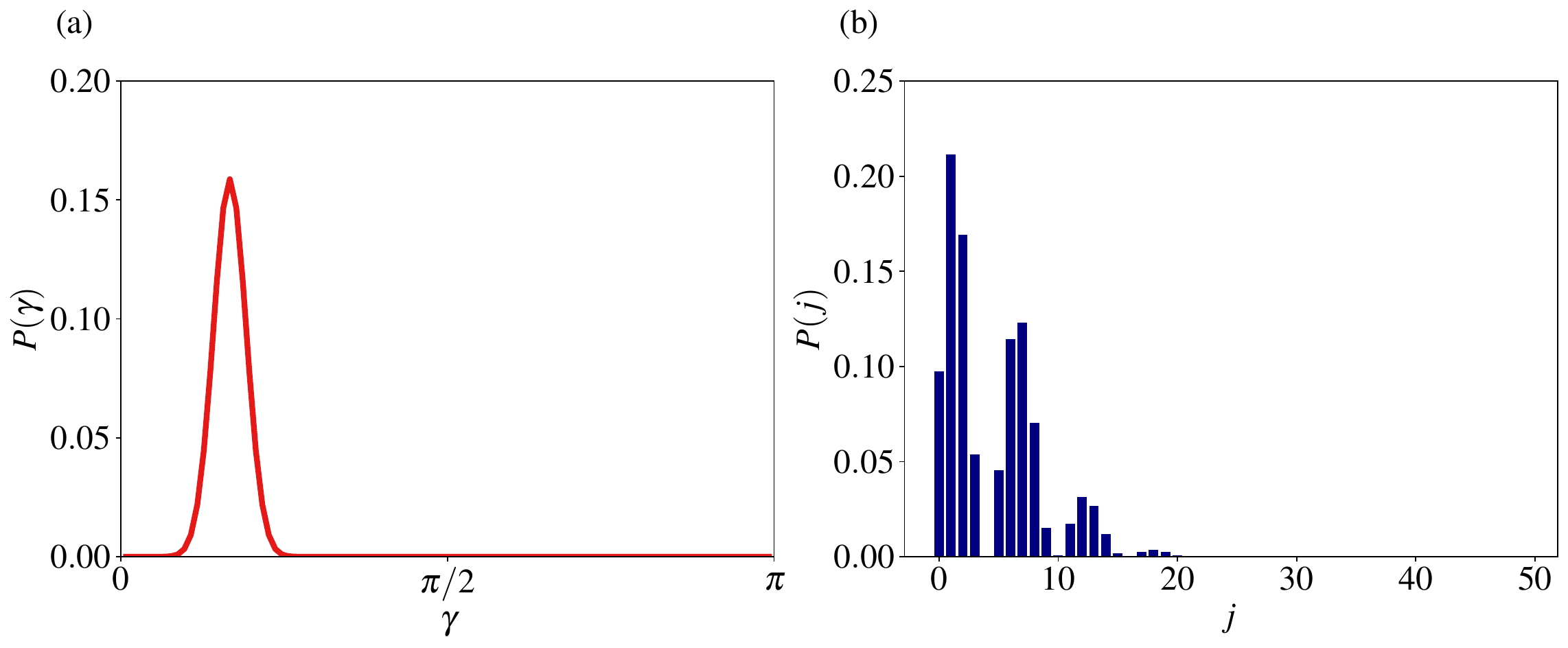}
    \caption{(a) Spatial distribution as a function of Jacobi angle $\gamma$ for an initial wave packet with $\gamma_0 = \pi/6$ and $\sigma_\gamma = \pi/20$. (b) The corresponding quantum nuclear rotational angular momentum distribution for the same initial wave packet.}
    \label{fig:Pthetaj}
\end{figure}

The total (nuclear + electronic + spin) wavefunction can be written as:
\begin{equation}\label{eq:Psi}
   \bra{R,r,\gamma} \ket{\Psi} = \sum_i \bra{R,r,\gamma} \ket{\Psi_i} =  \sum_{i} \chi_i(R,r,\gamma) \ket{\psi_i}
\end{equation}
where $\chi_i$ is the nuclear wavefunction on electronic state $\ket{\psi_i}$ and the index $i$ runs over our three-state spin-diabatic basis $\{S,T_{+1},T_{-1}\}$. 
The initial nuclear wavefunction is expressed as,
\begin{equation}\label{eq:chi}
    \chi_{i}(R,r,\gamma,t=0) =   \varphi(R) \phi(r) u(\gamma), 
\end{equation}
\begin{equation}\label{eq:phi0}
    \varphi(R) =  \left(\frac {1} {\pi\sigma_R}\right)^{\frac 1 4}\exp(-\frac {(R-R_0)^2}{2\sigma_R^2}-iK_0 R).
\end{equation}
Here, the initial average position $R_0$, initial momentum $K_0$ and width $\sigma_R$ together define a localized wavepacket moving in the $R$ direction. 
Normally, for most scattering calculations, $\phi(r)$ and $u(\gamma)$ would be chosen to be the rovibrational eigenstate of the diatomic component of the system. However, in this case, we are interested in stereodynamics scattering and so we will initialize the system in the vibrational ground state $\phi_{\nu_0}$ of \ce{H2} with a specific angular distribution (i.e., a Gaussian distribution characterized by $\gamma_0$ and $\sigma_\gamma$),
\begin{equation}\label{eq:phiu}
    \phi(r) u(\gamma) = \phi_{\nu_0}(r) \left(\frac {1} {\pi\sigma_\gamma}\right)^{\frac 1 4}\exp(-\frac {(\gamma-\gamma_0)^2}{2\sigma_\gamma^2}),
\end{equation}
which allows us to simulate scattering process with different incident scattering angles. 
Note that our initial setup differs from conventional calculations where the initial wave function is single nuclear rotational angular momentum eigenstate. To illustrate this difference, in Fig.~\ref{fig:Pthetaj}, we have plotted both (left) the spatial angular distribution and (right) the quantum nuclear rotational angular momentum distribution for one particular initial wave function.
Although this initial setup may be difficult to achieve experimentally, this approach will yield key insight into spin-dependent stereodynamical effects in chemical reactions.\cite{footnote1} 

{\em Ab initio} SA-CASSCF/MRCI calculations were carried out so as to generate a full dimensional singlet-triplet ISC potential energy surfaces. 
For the state-averaged complete active space self-consistent field (SA-CASSCF) calculations, a state-average was employed with equality between three states: the lowest singlet state (${}^1A'$) plus the two lowest triplets (${}^3A''$ and  ${}^3A'$), all
in an aug-cc-pVDZ basis set.
The active space contained 8 electrons and 10 active orbitals (9 $a'$ orbitals and  1 $a''$ orbital). Following the SA-CASSCF calculation, an internally contracted MRCI calculation  was performed.
Finally, once the  MRCI wavefunctions were computed, we evaluated the spin-orbit coupling matrix elements using the full Breit-Pauli operator.

\begin{figure}[ht]
    \centering
    \includegraphics[width=\linewidth]{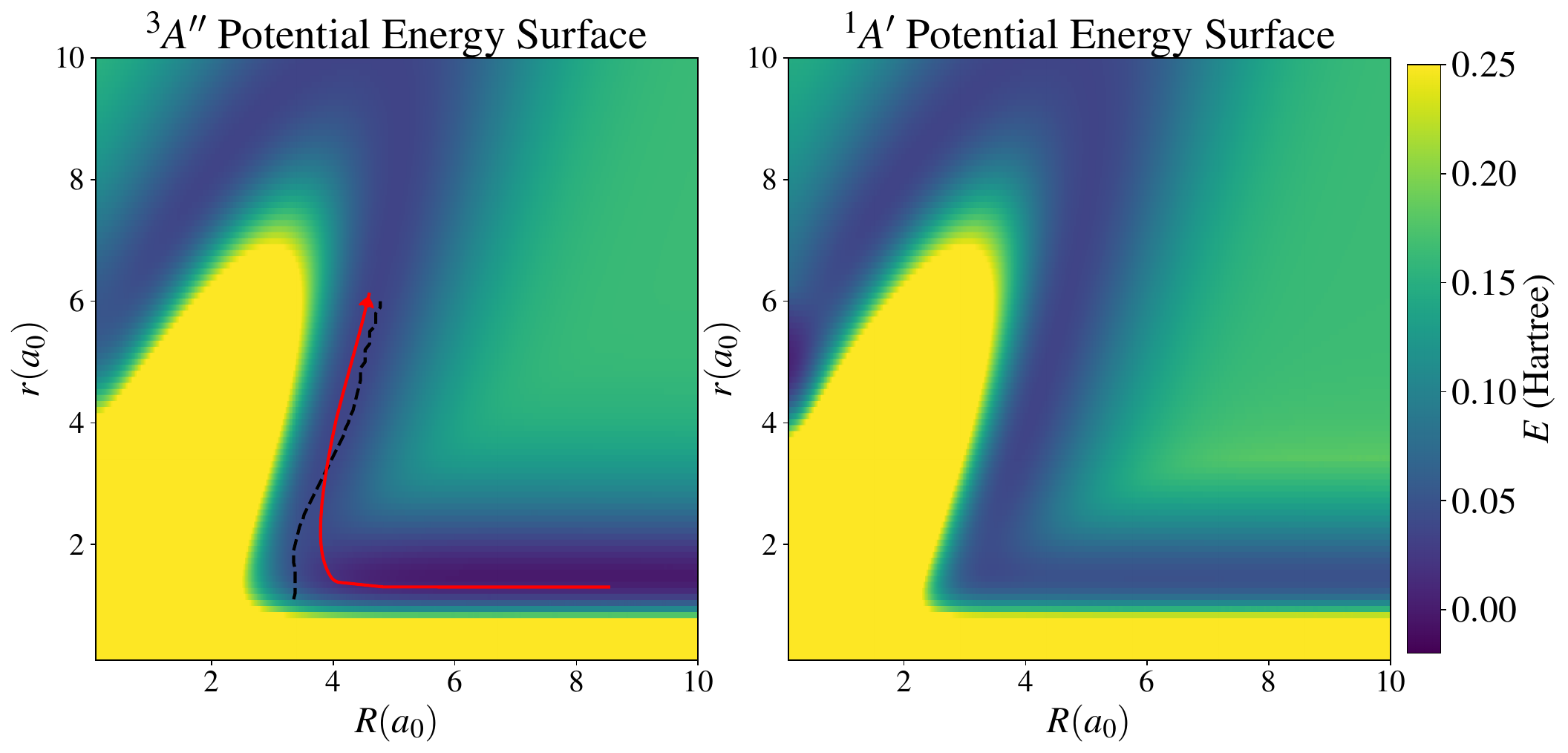}
    \caption{Colormap plot depicting the PES of ${}^3A''$ state for a constant Jacobi angle $\gamma = {\pi}/6$. The wavepacket will go through the singlet-triplet ISC seam (the black dashed curve) along the reaction path (the red curve with an arrow) from the \ce{S}/\ce{H2} basin to the \ce{SH}/\ce{H} basin.}
    \label{fig:PES}
\end{figure}

All electronic structure calculations were performed with the MOLPRO package \cite{werner2020}. To build a potential energy surface, we chose an evenly spaced grid for the $R$ and $r$ coordinates, $0.1a_0\le R \le 10.0 a_0$, $0.1a_0 \le r \le 10.0 a_0$,
with $N_R = 400$ and $N_r = 100$. 
For the coordinate $\gamma \in [0,\pi]$ we chose the discretization as follows. First, we sampled $\cos(\gamma)$ from a standard set of Gauss–Legendre quadrature points on $[-1,1]$ for $N_\gamma = 100$, $\left\{x_1, x_2, \ldots, x_{N_{\gamma}} \right\}$; second, we inverted these points with the cosine function, $\gamma_j = \cos^{-1} (x_j)$. 
These points sampled above are used directly as grid points for propagating our quantum wave packet dynamics below and no interpolation is performed.
In Fig.~\ref{fig:PES}, we show a plot of the ${}^3A''$ PES for a fixed angle $\gamma = \pi/6$. All energies are relative to the reactant asymptotic energy.

In Fig.~\ref{fig:PES}, we also show a simple low-energy path in configuration space from the \ce{S}/\ce{H2} basin to the \ce{SH}/\ce{H} basin; the singlet-triplet ISC seam is also shown. 
The nuclear wavepacket is initialized when the \ce{S} atom and \ce{H2} molecule are far apart (large $R$) on the triplet electronic PES, and one can imagine the  \ce{S} atom being scattered towards  \ce{H2} along  the red curve in Fig.~\ref{fig:PES}). Following the path in red, a nuclear wavepacket passes through the singlet-triplet ISC seam (the black dashed curve).

As mentioned above, we enforced zero total nuclear angular momentum for our simulations. 
At time zero, the nuclear wavepackets are initialized on one of the pure triplet spin-diabatic states $T_{+1}$ or $T_{-1}$ at $R_0 = 6.0a_0$ with $\sigma_R = 0.5a_0$. The initial momentum was set to be $K_0 = -25.0$ a.u (see Eq.~\ref{eq:phi0}). Dynamical calculations were run for different $\gamma_0$ values (see Eq.~\ref{eq:phiu}) but the angular distribution had a fixed width, $\sigma_\gamma = \pi/20$. 
For more details of our  three-dimensional wavepacket calculations, please see the supporting information.   

\begin{figure}[ht]
    \centering
    \includegraphics[scale=0.5]{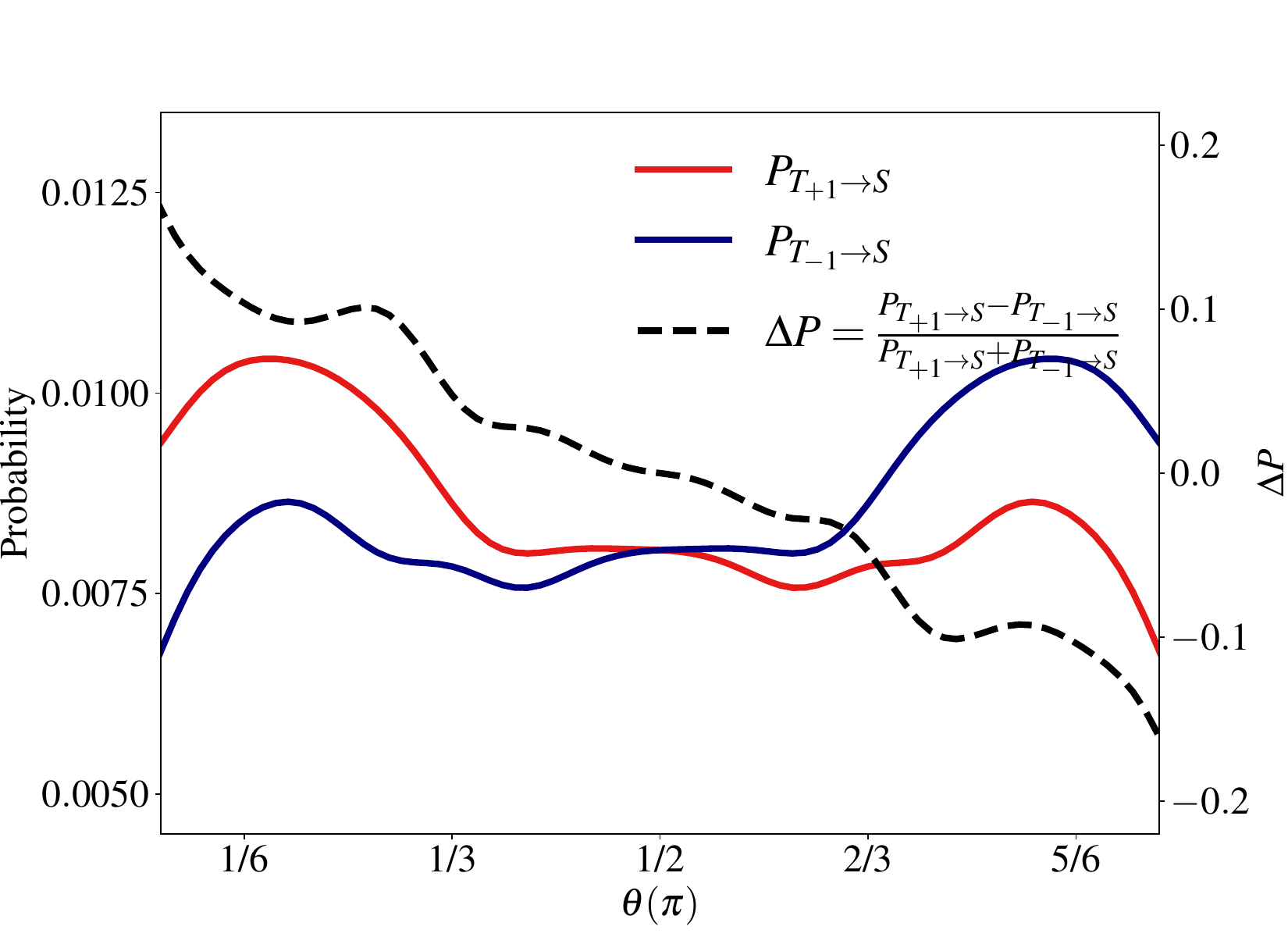}
    \caption{
    Reaction probabilities from triplets $T_{+1}$ and $T_{-1}$ to the singlet $S$ with an incident energy of 0.1 Hartree are plotted as a function of  incident Jacobi angles $\gamma$, demonstrating steoredynamical effects. 
    Furthermore, distinct ISC reaction probabilities emerge when systems are initialized with different spin-polarization at the same scattering angle.
    }
    \label{fig:Prob}
\end{figure}

\begin{figure}[ht]
    \centering
    \includegraphics[scale=0.75]{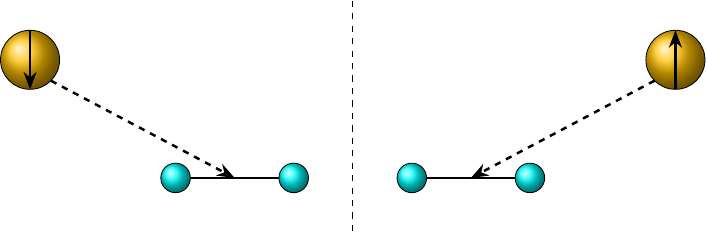}
    \caption {Upon a mirror reflection of the total system as shown in the figure, the incident Jacobi angle changes from $\gamma$ to $\pi - \gamma$ and the electronic spin is reversed. The ISC dynamics will be identical before and after this transformation.
    }
    \label{fig:mirror}
\end{figure}

To compute the final reaction probability, we calculated the cumulative probability flux for a dividing surface at $r_{s} = 5a_0$,
\begin{equation}
    P_i = \frac {\hbar} {\mu_{\ce{H2}}}\int_0^{+\infty}dt \Im(\bra{\Psi_i(t)}\delta(r=r_{s}) \nabla_r \ket{\Psi_i(t)}).
\end{equation}
Here, we evaluated $\nabla_r \ket{\Psi}$ with a fast Fourier transformation. 
In Fig.~\ref{fig:Prob}, we plot the cumulative reaction probability $P_{T_{\pm 1} \to S}$ from state $T_{+1}$ or $T_{+1}$ to state $S$ as well as the relative ``spin-polarized probability''   $\Delta P = \frac {P_{T_{+1} \to S} - P_{T_{-1} \to S}}   {P_{T_{+1} \to S} + P_{T_{-1} \to S}} $  for various incident scattering angles in the BF frame. We find that the ISC reaction probabilities clearly exhibit stereodynamical effects that are dependent on the incident scattering angle and the initial spin state. Note that even though the total probability for a triplet to transform to a singlet is small, the relative difference between $P_{T_{+1} \to S}$ and $P_{T_{-1} \to S}$ can be as large as $15\%$ for small incident angles.  
Note also that the reaction probabilities for two spin polarization show a symmetry around $\gamma = \pi/2$; more precisely, $P_{T_{+1} \to S}(\gamma) = P_{T_{-1} \to S}(\pi - \gamma)$.  
As show in Fig.~\ref{fig:mirror}, this symmetry can be easily understood by considering a mirror reflection along the axis  perpendicular to the diatomic bond with a corresponding reversal of spin-state. 
By considering the scenarios in Fig.~\ref{fig:mirror} from different viewpoints, one must conclude that both scenarios lead to the same dynamics and reaction probabilities. As a result, when the system begins with $C_{2v}$ symmetry, i.e., $\gamma = \pi /2$, it follows that the ISC reaction probabilities must be identical and there can be no spin-polarization -- as shown in Fig.~\ref{fig:Prob}.  Also, it is not hard to deduce that the spin-dependent reaction probabilities $P_{T_{+1} \to S}$ and $P_{T_{-1} \to S}$ for systems initialized with any of the diatomic rotational eigenstates will be the same. 

\begin{figure}[ht]
    \centering
    \includegraphics[width=\linewidth]{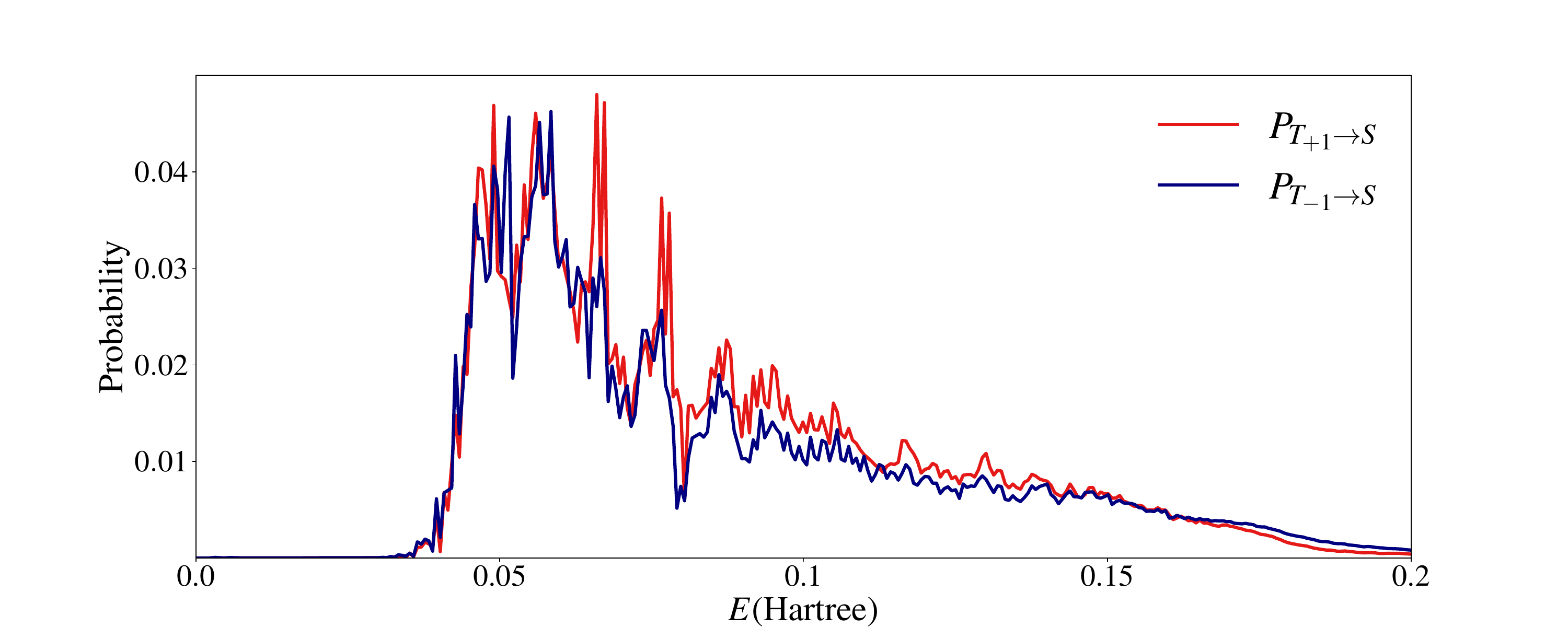}
    \caption{Energy-resolved reaction probabilities for systems initialized on $T_{+1}$ and $T_{-1}$ with incident scattering angle $\gamma =  \pi / 6$.}
    \label{fig:PE}
\end{figure}

Another quantity of interest is the energy-resolved reaction probability $P_i(E)$ \cite{Zhang1994},
\begin{equation}\label{eq:PiE}
    P_i(E) = \frac {\hbar} {\mu_{\ce{H2}}} \Im(\bra{\Psi_i(E)}\delta(r=r_{s}) \nabla_r \ket{\Psi_i(E)}),
\end{equation}
\begin{equation}\label{eq:psiE}
    \ket{\Psi_i(E)} = \frac 1 {a(E)} \int_0^{+\infty} dt e^{iEt} \ket{\Psi_i(t)}
\end{equation}
where $\ket{\Psi_i(E)}$ is calculated from a Fourier transform and $a(E)$ is the energy $E$ component of the initial nuclear wavepacket. 
As illustrated in Fig. \ref{fig:PE}, the probability of the ISC reaction for an incident angle of $\gamma =\pi/6$ varies for the two different initial spin-polarization over a range of energies. However, the reaction probability for reaching a singlet when starting from state $T_{+1}$ is larger than when starting from state $T_{-1}$ for most energies, including the intermediate incident energy, $E \approx 0.1$ Hartree. Thus, the data in Fig.~\ref{fig:PE} is consistent with the data in Fig.~\ref{fig:Prob}. It is also prudent now to consider the extreme cases  of low and high incoming energy.  At low incident energy, $E <  0.04$ Hartree, the nuclear wavepacket does not have enough energy to pass the triplet barrier so the reaction probability is close to 0. At very high energy, the dynamics become more diabatic and the wavepacket tends to stay on the same triplet spin-diabat and the ISC reaction probability vanishes. 

In the present work, we have analyzed if and how the \ce{S + H2 -> SH + H} insertion reaction may lead to spin-dependent stereodynamical effects. Our results show a measurable spin preference ($\sim 10\%$) for the intersystem crossing reaction probabilities at certain geometries. Given the fact that singlet-triplet ISCs are ubiquitous and energetically accessible in numerous molecular systems, this finding should be of interest as spin-dependent effects may well exist for many other reactions (with much larger molecules).
However, before we dive into {\em ab initio} or experimental investigations of more complicated systems, several points are worth mentioning.

To begin with, we note first that although the scattering products must depend sensitively on the nuclear orientation (or Jacobi angle $\gamma$) of incidence, meaningful statistics can always be (and were) extracted by looking at the differences between results for different spin states; moreover, given the fact that initial reactants are non-polar, steering effects are not completely dominant here as far as the ISC channel.  Second, although there is a meaningful {\em relative} difference in the  spin-dependent reactivities, we found the total ISC probability is rather small (only $\sim 1\%$) on an absolute scale for the current theoretical scattering setup. 
To amplify the spin-dependent ISC probability, one should consider systems with larger SOC (e.g., molecules that contain heavier atoms).  More interestingly, one can also consider systems in the condensed phase with more scattering events; after all, excited triplets are more energetically favorable than excited singlets and, with more scattering events and a faster drive to equilibrate, one can suppose that triplet populations will increase (and one can investigate possible triplet spin polarization). 
For both of these scenarios, the size of the calculation will be larger, with more nuclear and electronic DoFs, and as such,  exact quantum dynamics will likely not  be viable and we will need to use a semiclassical framework.

\begin{figure}[ht]
    \centering
    \includegraphics[scale=0.5]{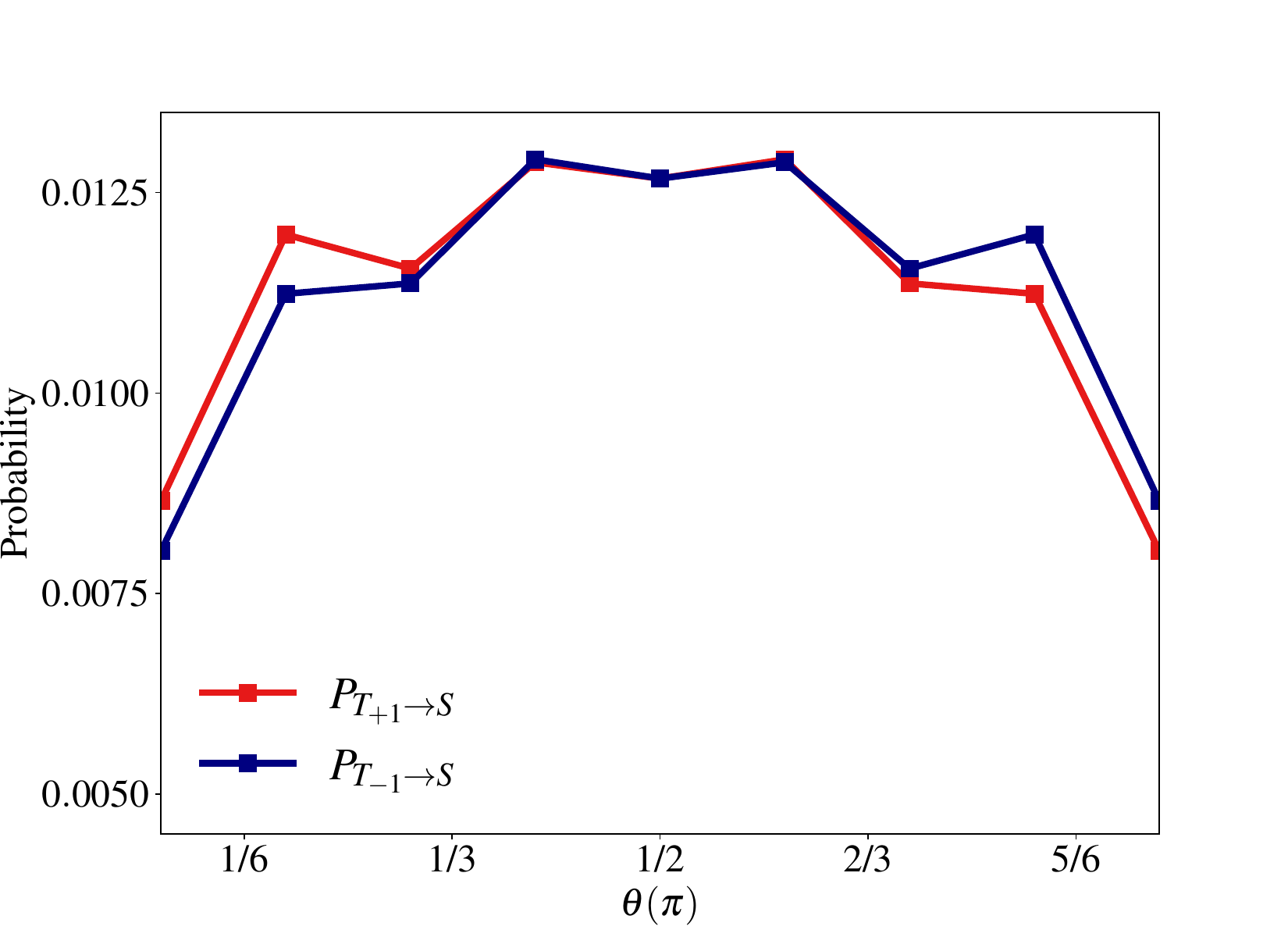}
    \caption{Reaction probabilities calculated from naive spin-diabatic surface hopping method.  
    }
    \label{fig:SHProb}
\end{figure}

To that end, we have explored the \ce{S + H2 -> SH + H} system above using a naive version of Tully's fewest switches surface hopping (FSSH) method. The FSSH simulations are propagated in a spin-diabatic basis with cubic-spline interpolated potential energy surfaces.
In Fig.~{\ref{fig:SHProb}},  we plot the ISC reaction probabilities calculated from FSSH method. For each data point,  $10^6$ trajectories are sampled from the same initial conditions as in Fig.~{\ref{fig:Prob}}.
We rescale the momentum along the direction of the difference of diabatic gradients $\vec \lambda_{jk} = \nabla E_j - \nabla E_k$ when trajectories hop between surface $j$ and $k$. We note that the results of ISC reaction probability are quite sensitive to the momentum rescaling directions (see below),
but overall, according to Fig. \ref{fig:SHProb}, naive FSSH gives results that qualitatively agree with the wavepacket results. 
We find a similar spin-polarized pattern as a function of incident scattering angle comparing to the quantum results,
which suggests that FSSH-like methods can serve as effective approaches for investigating spin-dependent non-adiabatic dynamics, which is a growing field of study nowadays.

To  understand how this spin asymmetry arises within the surface hopping, note that there is only one possibility: spin polarization emerges when the distinct coupling matrix elements in the electronic Hamiltonian described by Eq.~\ref{eq:ISC} lead to different quantum amplitudes. In other words, the different SOC matrix elements that couple $T_{+1}$ and $T_{-1}$ to $S$  will lead to different hopping probabilities between these spin states and, as Fig.~\ref{fig:SHProb} shows, these probabilities can be reasonably accurate. This finding is of particular interest because spin-diabatic FSSH does not conserve the total angular momentum for several reasons: (i) Hopping between states often produces changes in angular momentum depending on the rescaling direction.\cite{Shu2020}
As a practical note, we mention that if we employed a different momentum rescaling direction, e.g.  $\vec \lambda_{jk} = \Re\left(\nabla H_{jk}\right)$ and $\vec \lambda_{jk} = \Im\left(\nabla H_{jk}\right)$, our results were far worse compared to exact quantum calculations. Interestingly, here we have rescaled momenta in the direction of the gradient of the different in diabatic energies, $\nabla E_j - \nabla E_k$. Moreover, for any closed isotropic system, it is very easy to show that
${\bm R} \times {\bm F}_j = 0$. Thus, it follows that:
\begin{equation}
    \frac {d \bm J_{nuc}} {dt} = {\bm R} \times {\bm F}_j = 0 
\end{equation}
Thus, the present algorithm does in fact conserve nuclear angular momenta, which might partially explain our good results here. 

(ii) A second and more serious obstacle is that whenever runs classical Born-Oppenheimer dynamics along an isotropic surface,  one conserves only the {\em nuclear} angular momentum and not the {\em total}  nuclear plus spin plus  electronic (as computed by the amplitudes) angular momentum, which 
must cast some doubt on running spin-diabatic FSSH dynamics.
Now,  one usually performs FSSH in an adiabatic (rather than diabatic) basis \cite{Tully1998} and over the last few years, there has been a push to understand how to propagate adiabatic FSSH dynamics with spin DoF. One key ingredient arises in such a  case: nuclear Berry curvature effects\cite{Miao2019,Takatsuka2017}.  When running classical dynamics, such nuclear Berry curvature effects induce a pseudomagnetic field for the nuclear motion that in turn alters trajectories, leading to  angular momentum  transfer between nuclei and electronic components 
and ultimately enforcing {\em total} angular momentum conservation. 
\cite{Bian2023}.
These effects do not arise within spin-diabatic FSSH calculations, where the diabatic potential energy surfaces of $T_{+1}$ and $T_{-1}$ are identical and one always finds identical nuclear motion along both.  Incorporating nuclear Berry curvature effects within semiclassical dynamics methods (especially FSSH) is a nontrivial task, and developing meaningful algorithms for doing so will be an important step towards a quantitative understanding of spin-dependent non-adiabatic dynamics in more general/complicated systems \cite{Bian2022,Bian2022:PS}.

Lastly, let us go beyond the FSSH framework and consider the origin of 
this spin-dependency from a more general physical perspective. Quite generally, generating spin-dependent reaction products requires two ingredients -- time-reversal symmetry breaking and spatial inversion symmetry breaking.\cite{Zollner2020}
The need for time-reversal symmetry breaking can be understood from first-order perturbation theory. At equilibrium, the ISC Fermi Golden Rule rates $k_{FGR} = {2\pi}/{\hbar} \abs{\bra{T_{\pm 1}} H_{ISC} \ket{S}}^2$ are identical for the two triplet spin-diabats, which implies that introducing spin polarization must be a non-equilibrium property. 
In our calculations, the system exhibits strong non-equilibrium features by construction: we are running scattering calculations. It  will be very interesting to assess if/how spin polarization emerges in the condensed phase, e.g. in electron transfer problems, where equilibration occurs more quickly and the environment must be considered very carefully.
Next, the need for spatial inversion symmetry breaking can be analogously understood in the context of condensed matter theory.  Periodic electronic structure calculations with Rashba \cite{Rashba1960} and Dresselhaus \cite{Dresselhaus1955} coupling make clear that the spin components of electronic bands can split apart for systems without spatial inversion symmetry.  
As far as our present calculations are concerned, we recovered spin polarization only when we resolved   the scattering event over different incident Jacobi angles $\gamma$; if we were to average over $\gamma$, and assume an isotropic distribution of incoming angles, we would not see any spin polarization. Thus, if an experimentalists aims to see the ISC spin polarization we observe here for a scattering experiment, it will be 
crucial to align all the molecules in the laboratory frame, which can presumably be achieved by employing laser techniques in ultracold experiments \cite{Rost1992,Mukherjee2014} or attaching molecules to a surface \cite{Yoder2010}.

As a side note, before concluding, one more point about the \ce{S + H2} system is appropriate.  The presented three-dimensional quantum dynamics simulations can still be improved it at least {\bf two} aspects: (i) Here, we have considered only systems with a zero total nuclear angular momentum. More generally, for systems with non-zero total nuclear angular momentum, one must include four nuclear DoFs (the three included above and the projection of total nuclear angular momentum in the body-fixed $z$-axis $J^z_{nuc}$) and consider Coriolis couplings between different $J^z_{nuc}$ states.
(ii)  Our dynamics here go beyond electrostatic Born-Oppenheimer calculations and the presence of spin-orbit coupling changes the electronic spin angular momentum for a system with both singlet and triplet character. Thus, if one wished to describe the \ce{S + H2} dynamics exactly, one would need formally to include as a degree of freedom the total {\em nuclear} angular momentum ${\bm J}_{nuc}$ (which is not a good quantum number for systems with  SOC). Implementing such five dimensional simulations and exploring the subsequent  molecular scattering dynamics remains another  future research direction if one is interested in the most accurate possible bimolecular scattering reactions.

In summary, we have discussed how spin-dependent reaction probabilities can and do emerge for a well-known ISC problem in chemical physics.
We find the three ingredients are necessary for realizing spin-dependent chemical reactivity: spin-orbit coupling, time-reversal symmetry breaking and spatial inversion symmetry breaking.  Interestingly, naive spin-diabatic mixed quantum classical FSSH calculations are qualitatively accurate for the dynamics investigated here, but as mentioned above, that may be coincidental: the FSSH calculations here do conserve nuclear angular momentum, but mostly ignore the transfer of angular momentum between electronic and nuclear degrees of freedom.

Looking forward, the subject of angular momentum transfer and conservation are directly relevant to the CISS phenomenon and the present results suggest semiclassical approaches should be helpful studying larger systems (with more than three atoms) in the future.  In particular, given the significance of nuclear motion in the incoherent charge transfer mechanism observed in organic molecules like DNA \cite{Xiang2015,Kim2016} where a CISS effect is observed, as well as the temperature dependence observed within CISS experiments \cite{Das2022,Alwan2023}, there is a very strong reason to further develop mixed-quantum classical calculations that do treat spins, electrons, and nuclei consistently. Since the present calculations clearly reveal measurable spin-dependent effects for small systems, one wonders if simulations on larger systems will reveal larger effects, potentially  helping to understand and explain CISS.

\section{Acknowledgments}
This work has been supported by the U.S. Department of Energy, Office of Science, Office of Basic Energy Sciences, under Award no. DE-SC0019397 (J.E.S.). X.B. thanks Yanze Wu for numerous helpful discussions.

\begin{suppinfo}
Additional details of our wave packet dynamics simulations and some one dimensional cuts of the potential energy surface can be found in the supporting information. 
The source code of our potential energy surface is available on GitHub at \url{https://github.com/xzbian/H2S}.
\end{suppinfo}
  
\bibliography{main.bib}

\begin{mcitethebibliography}{54}
\bibitem[foo()]{footnote1}
As a practical matter, there is a nuance which may make this setup difficult to
  achieve. Note that, because we have assumed that the total nuclear angular
  momentum $J=0$ in Eq.~3, the distribution of nuclear rotational angular
  momentum $j$ sampled in Fig.~2 must be compensated by the nuclear orbital
  angular momentum of the atom-diatom system in order to apply the theory
  stemming from Eq.~3, and achieving such a superposition of initial states may
  be difficult in practice. In other words, the resulting nuclear wave
  functions from Eq.~7 are not equivalent to a diatomic pendular state, which
  is usually constructed as a superposition of rotational angular momentum
  eigenstates with different total nuclear angular momenta.\relax
\mciteBstWouldAddEndPunctfalse
\mciteSetBstMidEndSepPunct{\mcitedefaultmidpunct}
{}{\mcitedefaultseppunct}\relax
\EndOfBibitem
\end{mcitethebibliography}
\end{document}